\documentclass[aps,chaos,twocolumn,showpacs,preprintnumbers,amsmath,amssymb,groupedaddress,superscriptaddress]{revtex4-1}

\usepackage{graphicx}
\usepackage{dcolumn}
\usepackage{bm}
\usepackage{subfigure}

\usepackage{amsmath}
\usepackage{amsthm}
\usepackage{amssymb}

\begin{document}

\date{April 25th, 2018}

\title{Critically Slow Learning in Flashcard Learning Models}

\author{Joel Nishimura}
\affiliation{School of Mathematical and Natural Sciences, Arizona State University, Glendale, AZ 85306, USA}

\begin{abstract}

Algorithmic education theory examines, among other things, the trade-off between reviewing old material and studying new material: time spent learning the new comes at the expense of reviewing and solidifying one's understanding of the old. This trade-off is captured in the `Slow Flashcard System' (SFS) -- a system that has been studied not only for its applications in educational software but also for its critical properties; it is a simple discrete deterministic system capable of remarkable complexity, with standing conjectures regarding its longterm behavior. Here we introduce a probabilistic model of SFS and further derive a continuous time, continuous space PDE model. These two models of SFS shed light on the longterm behavior of SFS and open new avenues of research.

\end{abstract}

%

\maketitle

\section{Introduction}
That similar events recur is the foundation for both the value of  learning and its feasibility. After all, if every experience was entirely exceptional there would be little to learn and it would be of little value.  Thus, at least abstractly, understanding a learning process is connected to understanding the recurrence and timing of events on which learning occurs.  
For instance, psychologists have thoroughly studied the `spacing effect,' where spacing reviews over time yields superior retention than a single more intensive review \cite{ebbinghaus1913memory,dempster1988spacing,cepeda2006distributed}.

More generally, when a learner is tasked with learning many things, any sequence of events or items exhibits some trade-off between the familiar, which reinforces understanding, and the new, which expands it.  While at once abstract, this trade-off is now also an explicit dilemma for the designers of educational software, such as Duolingo \cite{Duolingo}, who must determine the sequence in which previous content is reviewed and new content is introduced.

These real-world software applications have motivated recent mathematical models of this trade-off. Following the lead of \cite{novikoff2012education}, the sequence of educational units is frequently modeled in terms of a deck (or decks) of `flashcards', where a student repeatedly draws flashcards to review and the rules governing how cards are re-inserted in the deck determines the sequence of educational units.  These models and later adaptations have been used to address both the practical applications of designing educational software systems \cite{reddy2016unbounded,reddyaccelerating}, theoretical questions \cite{rodi2015optimal,lewis2014combinatorial} and optimizing the learning of a neural network \cite{amiri2017repeat}.  For instance, when there are relationships between the content to be reviewed, such as words with similar Latin roots, the related structure of content can be used to optimize the flashcard sequence\cite{rodi2015optimal}. Similarly, the large datasets of online platforms can be used to fit models of the spacing effect, and these can be used to determine the optimal draw sequence from a Leitner system of flashcards \cite{reddy2016unbounded}, where misremembered content is more frequently reviewed \cite{Leitner}.  

Despite these recent advances, there remains fundamental unanswered questions    regarding the trade-off between between the introduction of new material and the review of old material \cite{novikoff2012education}. On one hand, there is a clear distinction between the  models that allow for an infinite amount of material to be covered and remembered, and those models in which learning eventually stalls \cite{novikoff2012education}. On the other hand, the complex behavior of models at the critical point, where an infinite amount of material is covered, but new material is covered only `slowly' is not well understood. Indeed, the behavior at such critical points is often both richly interesting and indicative of the behavior in the complex systems that self-organize towards such critical points. As described in \cite{novikoff2012education}, the behavior of this critical point in learning is exemplified by the `Slow Flashcard System' (SFS) described below.

\begin{figure}
\includegraphics[width=\linewidth]{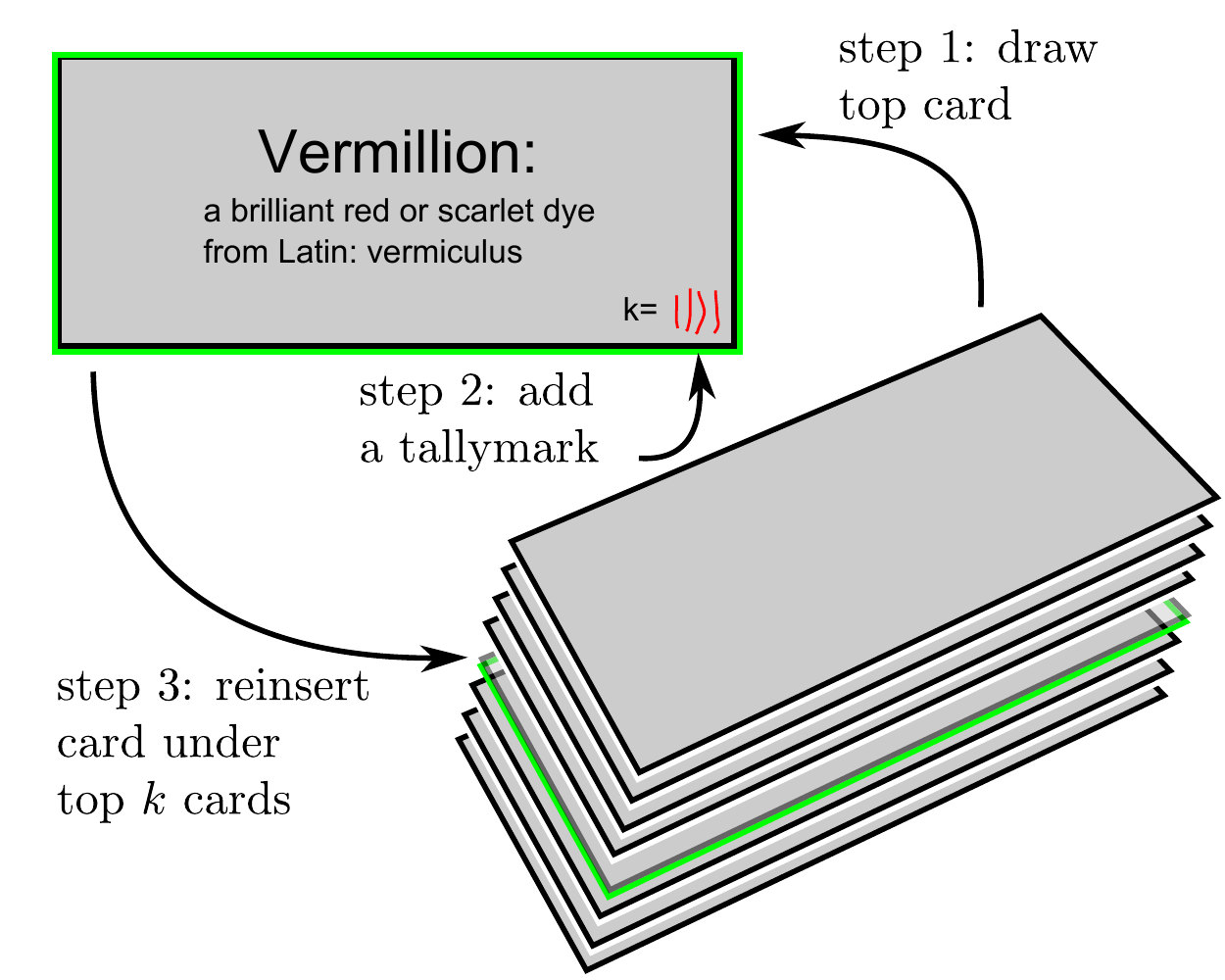} 
\caption{  The SFS consists of 3 steps.  
} \label{schematic}
\end{figure}

	The Slow Flashcard System has simple rules. Imagine a deck of vocabulary flashcards, as in Fig. \ref{schematic}, where on the corner of each card are tally marks denoting the number of times that that card has been viewed (i.e. studied or reviewed) by a student. At each time step, a card is taken off the top of the deck, a tally mark is added to the corner, and the card is reinserted into the deck underneath as many cards as there are tally marks on the card (including the newly added one).  For instance, if a card is drawn with $k$ tally marks then a $(k+1)^\emph{st}$ mark is added, and the card is then reinserted into the deck underneath the top $k+1$ cards, into position $k+2$. 

If the flashcards in a deck are consecutively labeled $a, b, c, \dots $ and all cards start with zero tally marks, then drawing cards and reinserting them will lead to a sequence: $a,b,a,b,c,a,c,b,d,c,d,a,b,d,c$.  Over time, new as-yet-unseen cards naturally make their way into the sequence, emulating how a learning process incorporates new material.  

	The Slow Flashcard System is deterministic and yet produces a combinatorially complex sequence that some have compared to a psuedo-random number generator. In order to gain insight into this deterministic system, we model it with a discrete time probabilistic system and then take the continuous time limit.

	In \cite{novikoff2012education} it is proven that starting with a deck of cards without any tally marks, the sequence has the property that after the $k^\emph{th}$ occurrence of any card, the number of time steps before the $(k+1)^\emph{st}$ occurrence will always be between $k$ and $k^2$. In \cite{lewis2014combinatorial} these bounds are tightened to $k$ and $2k$, settling a conjecture from \cite{novikoff2012education}. 

\begin{figure}
\includegraphics[width=1\linewidth]{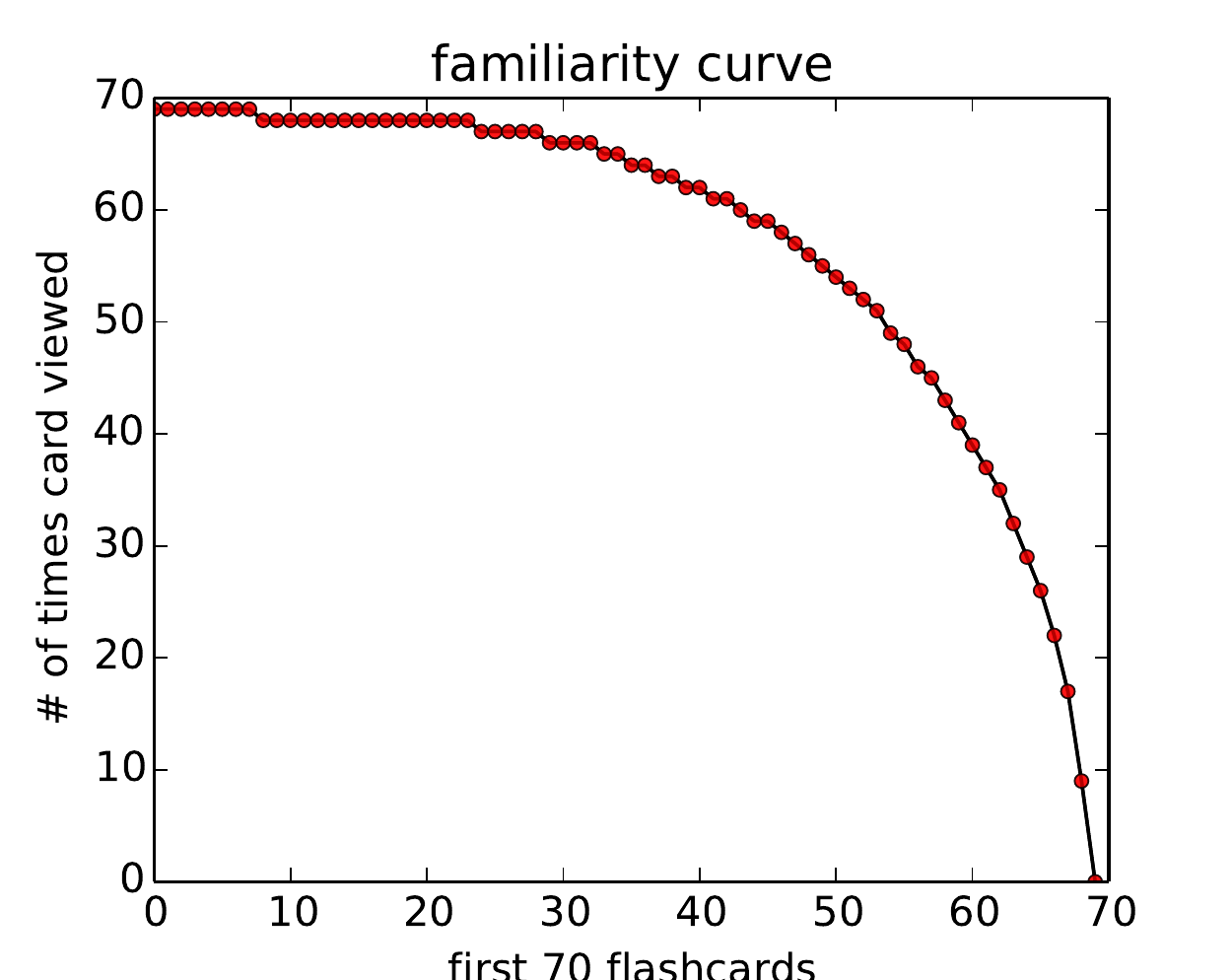} 
\caption{  After drawing $4000$ cards, $69$ unique cards have been viewed at least once, and the very first card has been viewed $69$ times.  The relationship between the original depth of a card and the number of times that card has been viewed has a robust and consistent shape in the SFS.
} \label{firstFamiliarity}
\end{figure}

	The Slow Flashcard System is conjectured in \cite{lewis2014combinatorial} to have another interesting feature, which we will study in this paper. Consider Fig. \ref{firstFamiliarity}, which plots the number of times each flashcard has been seen after $4000$ total flashcards have been viewed.  The shape of the curve in  Fig. \ref{firstFamiliarity} describes the balance between new material that has barely been seen and old material that has been viewed many times. Moreover, while the number of cards that have been viewed can only increase, the shape of the relationship in Fig. \ref{firstFamiliarity} appears robust, suggesting that there exists a `familiarity curve' that describes the limiting relationship. The existence of a familiarity curve -- the notion that the plots actually approach a fixed curve at all -- was first conjectured in \cite{lewis2014combinatorial}, though in the language of tableaus.  It was shown in \cite{lewis2014combinatorial} that if the familiarity curve exists it must lie below a circle $x^2+y^2 = 2+\epsilon$ and above the line $x+y=1$. However, thus far there has been only numerical evidence of the curve's existence. In this paper we develop a PDE that describes the observed behavior of the familiarity curve. 

Moreover, our analysis elucidates another mystery: why the familiarity curve appears not to be specific to the Slow Flashcard System; it seems to arise in a broad class of variants of SFS as well \cite{lewis2014combinatorial}. 

\begin{figure*}
\includegraphics[width=1\linewidth]{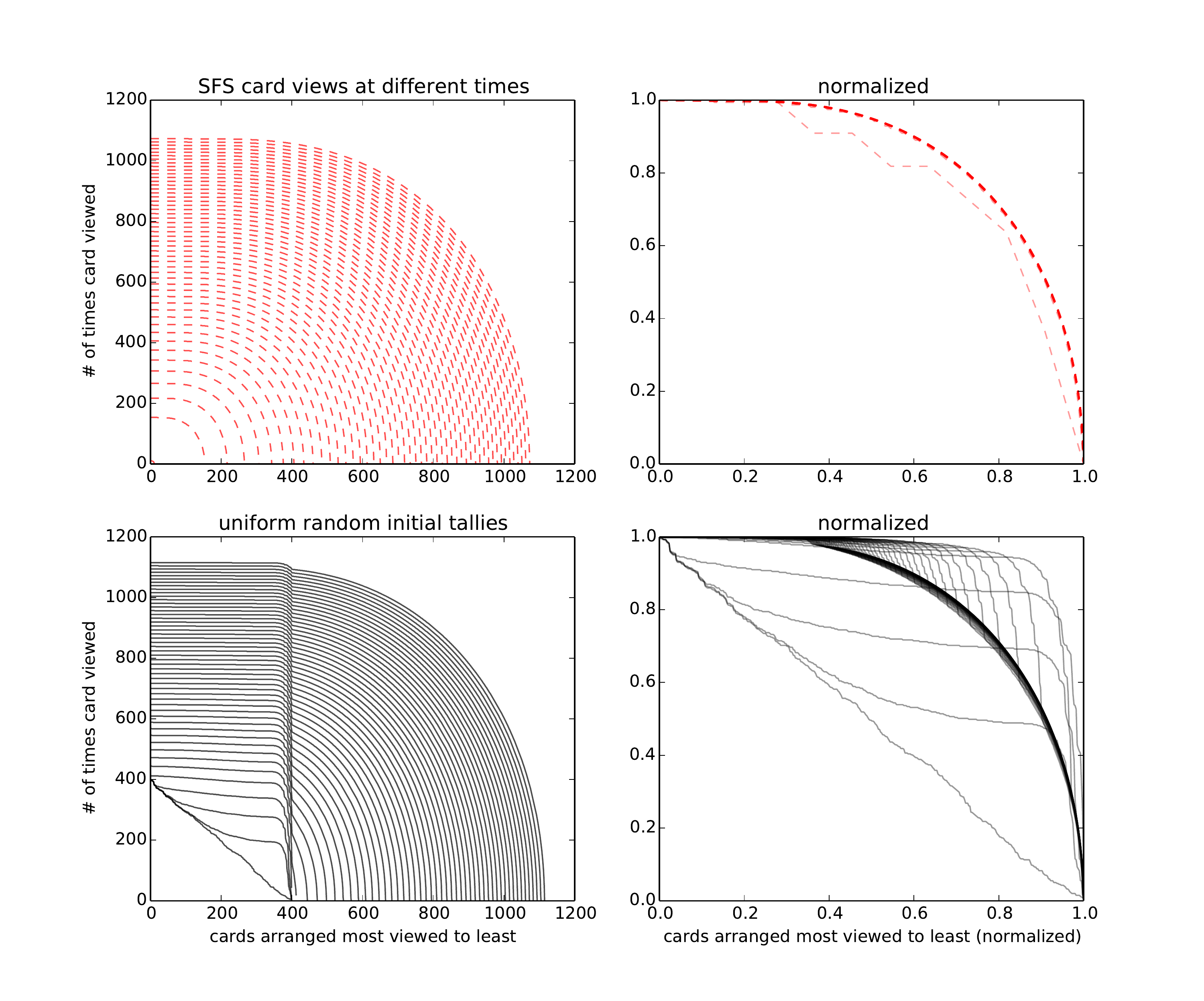} 
\caption{  To visualize the state of a deck of flashcards, at a time $t$ arrange the cards in order of decreasing tallymarks and plot these cards' decreasing number of tallymarks (left). Notice that these curves have the same shape, in that rescaling them to fit inside the unit square collapses them to the conjectured familiarity curve (right). The familiarity curve appears to represent the long term limiting behavior of a variety of initial decks, whether the deck begins with a deck of cards with a random number of tally marks (bottom) or an initially blank set of cards (top).   
} \label{RandomDeck}
\end{figure*}

In the next section, \emph{The Familiarity Curve}, we describe the curve in more detail and show the results of some numerical simulations. In \emph{The Probabilistic SFS} we describe a new model which captures key features of SFS, but which allows for an analysis in the spirit of statistical mechanics rather than the combinatorial approach taken in \cite{novikoff2012education,lewis2014combinatorial}. In \emph{Analysis of the Probabilistic SFS} we derive a PDE whose only static solution corresponds to the familiarity curve. Further analysis of the PDE leads to yet more conjectures about the deceptively simple SFS, which we state in the final section of this paper.

\section{The Familiarity Curve}

In the SFS, the number of tally marks that a card has is the number of times that  card has been reviewed, and presumably, a measure of the understanding of that card's content. By examining the distribution of tally marks on the cards, we can probe whether the student has a broad and shallow understanding, a deep and narrow understanding, or potentially some pattern of understanding in between. 

To visualize this, consider the curve created, as in Fig. \ref{RandomDeck}, by sorting the cards from most familiar (most tally marks) to least familiar (least tally marks) and plotting their level of ``familiarity'' (the number of tally marks on each card). 

Fig. \ref{RandomDeck} (top) shows this curve for a student who begins with no outside knowledge and then studies using the SFS. Alternatively, Fig. \ref{RandomDeck} (bottom) shows the familiarity curve for a student who begins the SFS with $400$ cards and each card with a number of tally marks chosen uniformly at random in $[0,400]$.  Notice that while two students may begin with initially quite different distributions, evolution under the SFS tends to drive them towards the same shapes. To capture the changing shape of these curves we rescale them to fit inside the unit square, as in Fig. \ref{RandomDeck}. (Note that only those cards with at least one tally mark are included; this step in the normalizations will play an important role in the development and analysis of the probabilistic model later on.) We call these rescaled curves `familiarity curves'. 

These numerical studies suggest that perhaps the SFS drives all initial card distributions towards the same familiarity curve. This might be very hard to prove for SFS, but the probabilistic model presented below sheds light on why it might be true. 

One can interpret the limiting familiarity curve as representing a sort of perpetually well rounded but still improving student: the plateau on the left represents a broad set of cards, or eductional units, that are very familiar to the student, while the steep slope on the right represents a small set of newly introduced cards that are being rapidly reinforced.

If, as the numerical evidences suggests, it is the case that the SFS evolution always drives card distribution to the same limiting familiarity curve, then it suggests that it is the ongoing learning process, more than the initial knowledge of students, that determines the long term characteristics of a student.  Moreover, the particular shape of a student's familiarity curve may offer clues about the particular learning process that the student has gone through.

\section{SFS: A Probabilistic Understanding}
The sequence of cards produced by SFS is complicated, and so the task of deriving the formula for the familiarity curve, or even proving its existence, is daunting.  Given this complexity, we approach the system as if it were random instead of deterministic, first establishing from basic assumptions the probability that a given card arrives at the top of the deck at a given time step, and then use that probability to derive a partial differential equation that describes the evolution of the SFS.

\subsection{The Probability of Drawing a Card} 
To build a probabilistic model of the SFS we begin with a simple observation.
From the basic rules that define the SFS, one knows that cards that have been seen $k > 0$ times must always be in position $k+1$ or lower. It turns out, this fact, and several uniformity assumptions, can be used to create a model that produces not only approximately the correct draw statistics, but also lends itself to somewhat complete statistical analysis. 

To illustrate our proposed model, consider the following example. Suppose that the deck in SFS has $2$ cards with two tally marks, and $1$ card with one tally mark. Then if we temporarily identify cards solely by their tally marks, the possible deck ordering are: $122$, $212$.  Notice that it wouldn't be possible to have deck $221$, with the 1 in the third position in the deck, because a card with $1$ tally mark could only ever be in the first or second position in the deck, according to the rules of SFS.  (More generally, a card with $k$ tally marks can only ever be found in position $k+1$ or lower.) Thus of the four possible valid decks (excluding the two that have the card with 1 tally mark in the third position), two have tally mark sequence $122$ and two have tally mark sequence $212$. Therfore, a deck picked uniformly at random from this set of four valid decks will have the top card (in position 1) with tally mark $1$ half the time and $2$ the other half.  

More generally, suppose that at a given time $t$ we have a function $f_t(k)$ describing the number of cards in the deck with $k$ tally marks. Given any distribution $f_t$, there is a set of possible valid decks with this distribution of tally marks. Thus if we consider a deck picked uniformly at random from this set of decks, we can in theory deduce the probability $P(k)$ that the top card has $k$ tally marks.

\begin{figure}
\includegraphics[width=.9\linewidth]{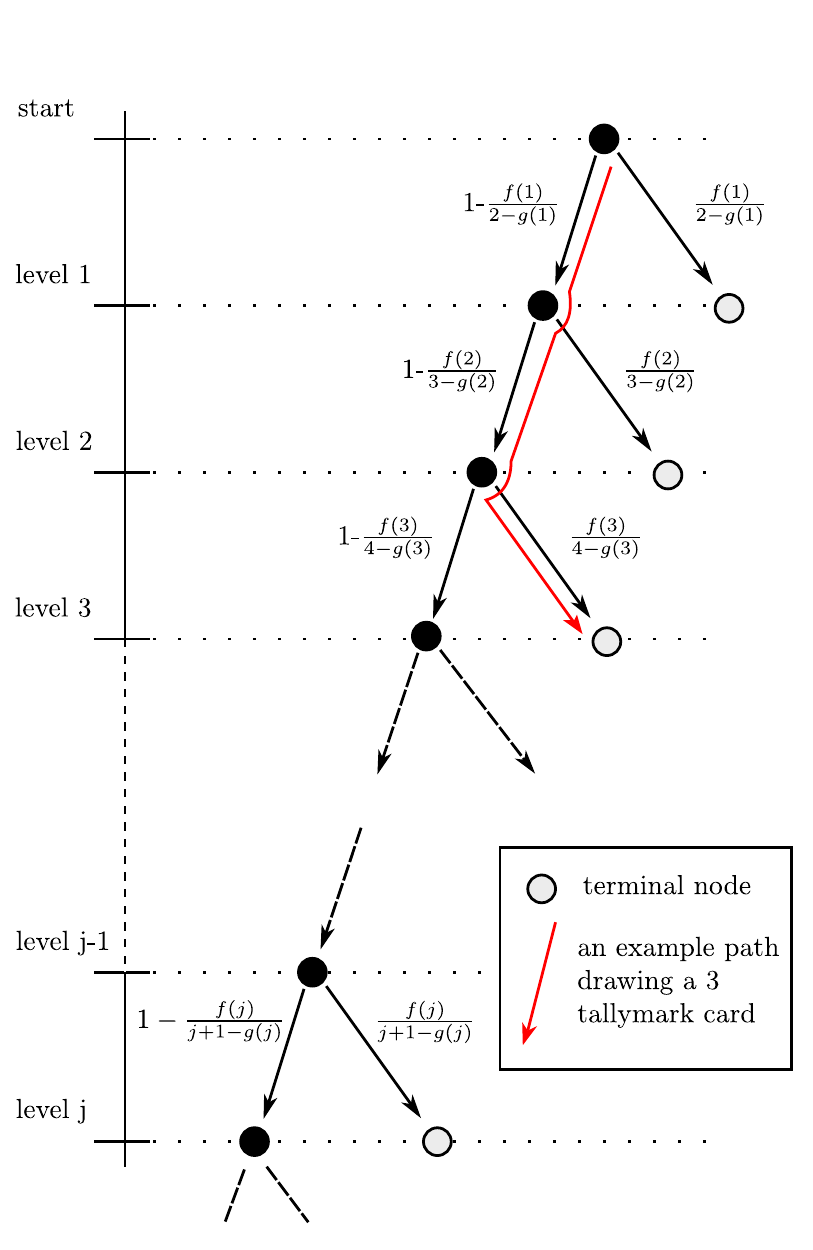} 
\caption{The probability of drawing a card with $k$ tally marks can be calculated from $f$ through a decision tree.} \label{DTree}
\end{figure}

Moreover, it's possible to calculate the probability without enumerating the possible decks, and we do so below. Once we have $P(k)$ in hand, we will be able to use that to describe how $f_t$ changes over time, and thus create our probabilistic system. We will observe at that point how the statistical behavior of this new system is extremely similar to SFS, and from the probabilistic system we will derive the PDE model.

Let us return to the example above, with $f(2)=2$ and $f(1)=1$.  What is the probability that the top card has $1$ tally mark? Since there is only $1$ card with $1$ tally mark, and that card can be in two possible spots, then there is $\frac{1}{2}$ probability the top card has $1$ tally mark. Conditioned on the top card not having $1$ tally mark, the probability that the next card has $2$ tally marks is $1$, since there are two cards with two tally marks and only $2$ places those cards can go (the $1$ tally mark card is taking up one of the $3$ possible positions cards with $2$ tally mark cards can go).  

To see how the interaction between cards extends, consider now the case where 
$f(1) = 0$, $f(2) = 1$, $f(3) = 2$ and $f(4)=2$.  The card with two tally marks has three positions it could fill, so the probability it is the top card is $\frac{1}{3}$.  If the top card does not have two tallymark then the second or third card must, and the deck must look like either $\_2\_\_\_$ or $\_ \_2\_\_$.  Conditioned on the top card not having two tally marks, there are three spots for the two cards with three tally marks, giving a conditional probability of $\frac{2}{3}$ and an unconditioned probability of $\frac{2}{3}(1-\frac{1}{3}) = \frac{4}{9}$.  If the top card does not have one, two or three tally marks then the deck is either $\_233\_$ or $\_ 323\_$, leaving $2$ spots for the remaining $2$ four tally mark cards, giving them a conditional probability of $1$ and an unconditioned probability of $1(1-\frac{2}{3})(1-\frac{1}{3}) = \frac{2}{9}$.

Extending this logic leads to a natural recursive formula. Conditioned on no cards with fewer tally marks being the top card, the probability of a card with $k$ tallymarks being drawn is 
\begin{equation*}
\frac{f(k)}{\text{number of available positions}}.
\end{equation*}
If there are $g(k)$ cards with less than $k$ tally marks 
($g(k)=\sum_{i<k} f(i)$), then the number of positions available is $k+1-g(k)$.
Since the probability that no cards with less than $k$ tally marks are on the top is $1-\sum_{i<k}P(i)$, the probability that the next card has k tally marks is simply
$$P(k) = \frac{f(k)}{k+1-g(k)}\left[ 1-\sum_{i<k}P(i) \right]\label{recursiveProb}.$$

This probability can also be seen as the output of the decision tree seen in Fig. \ref{DTree}, giving rise to alternate but equivalent expressions:
\begin{equation}
P(k) = \frac{f(k)}{k+1-g(k)}\Pi_{i<k} \left[ 1 -  \frac{f(i)}{i+1-g(i)},   \right] \label{recursiveProbOne}
\end{equation}
and,
\begin{equation}
\frac{P(k+1)}{P(k)} = \frac{f(k+1)}{f(k)}\frac{k+1-g(k+1)}{k+2-g(k+1)}. \label{recursiveProb}
\end{equation}

\subsection{The Probabilistic SFS}
In order to evaluate the accuracy of this probabilistic model we consider the behavior of a card system updated via this probabilistic rule. Consider a flashcard system where the state of the deck is not tracked, but instead, only the tally mark distribution $f$ is tracked, and each time a card is drawn, a card with $k$ tally marks is drawn with probability $P(k)$. The system size increases with the insertion of a card with $0$ tally marks each time the largest tally mark in the deck increases, by analogy with SFS, where the number of cards with non-zero tally marks periodically increases.

\begin{figure}
\includegraphics[width= \linewidth]{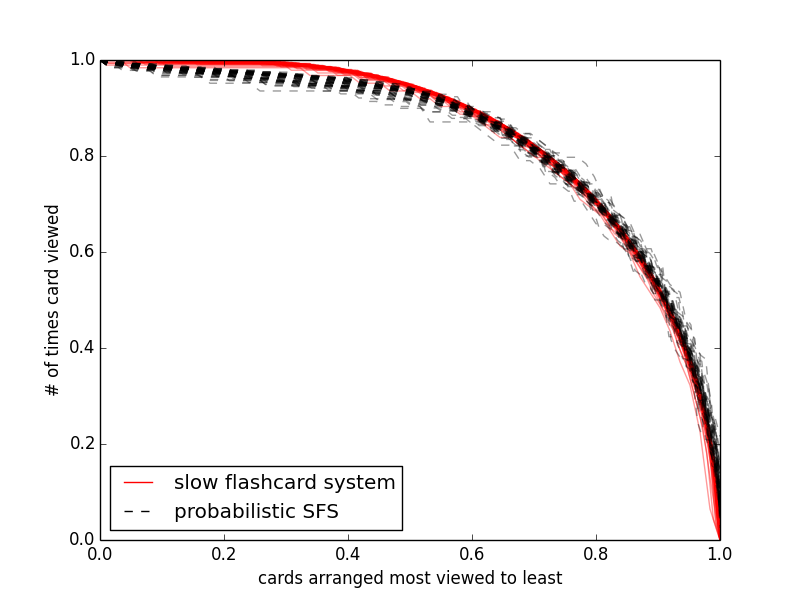} 

\includegraphics[width= \linewidth]{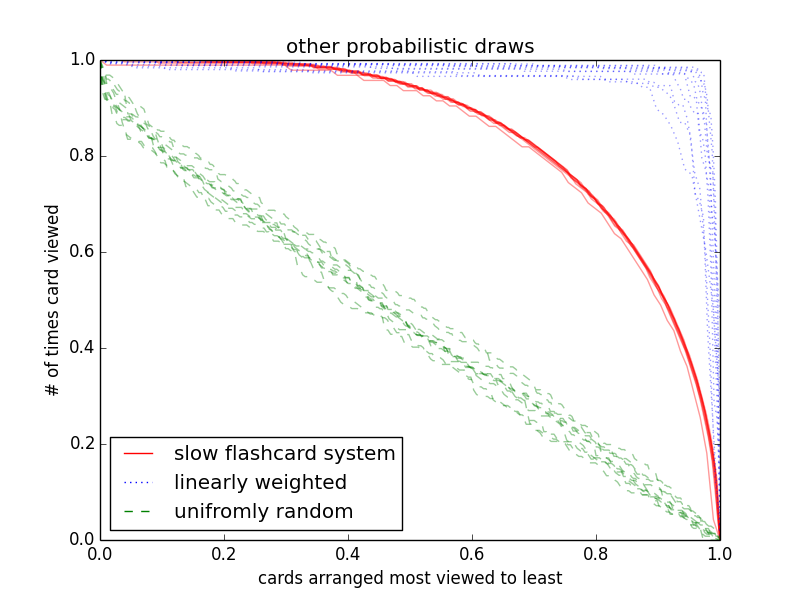} 
\caption{The familiarity curve created by the probabilistic SFS approximates the deterministic SFS (top), while using different probabilistic models, such as picking cards uniformly at random, or with probability proportional to their number of tallymarks does not approximate the SFS (bottom).} \label{probabilisticModel}
\end{figure}

As seen in Fig. \ref{probabilisticModel}, numerical simulations of the new probabilistic system produces decks with very similar statistics to the original deterministic SFS. In this way the probabilistic SFS suggests that the important feature of the original SFS wasn't the complexity of deck permutations, but the statistics of the tally mark weights and the simple property that cards with $k$ tally marks must be in positions less than or equal to $k+1$.    
Moreover, as we will see, this probabilistic system is conducive to analytic analysis and this analysis will predict the presence of an invariant familiarity curve.

\section{Analysis of the Probabilistic SFS}
While the SFS is particularly hard to analyze, the probabilistic SFS eliminates the permutation-based complexity that we have shown numerically to be unimportant to the overall statistics. We take advantage of this new probabilistic SFS to derive a continuous time version from it.

Suppose that we take a single step of the above probabilistic SFS. Since the number of tally marks on a card can only increase by one at any given time step, $f(k)$ will only increase if a card of level $k-1$ is promoted to level $k$ and will only decrease if one of level $k$ is promoted to level $k+1$. Let $\bar{f}$ be the expected value of $f$, then  to look at the expected change in tally marks, we see that: 
\begin{equation}
\bar{f}_{t+1}(k) = \bar{f}_t(k) + P_t(k-1)-P_t(k)  \label{DiscreteEqn}
\end{equation}

While this is still a discrete system, it is natural to consider if there is a limiting continuous case, where a continuous distribution $u$ could be studied rather than $f$.  Moreover, the natural domain of $u$ would be between $0$ and $1$, where $u(x)$ has argument $x=\mathrm{lim}_{t\to \infty} \frac{k_t}{n_t}$, which merely tracks how many tally-marks a card has compared to the card with the most tallymarks. However, before this difference statement can be used, we need to notice a peculiar feature of this system; namely, the system size may dynamically change from time $t$ to $t + 1$, so that eventually there are cards with arbitrarily many tally marks. Let the rate at which new cards are created be $\omega$; notice that omega is exactly the probability that the maximum tally mark in the deck increases. 

The changing system size means that the difference on the left: $\bar{f}_{t+1}(k) - \bar{f}_t(k)$, is actually a difference across both $t$ and the relative size of cards with $k$ tally marks.  As seen in the appendix, the appropriate scaling of time into $\tau$ and tally marks into $x$, transforms the difference $\bar{f}_{t+1}(k) - \bar{f}_t(k)$ into 
$u_\tau(x,\tau) - (\omega x)u_x(x,\tau)$
The multiplicative factor in front of $u_x$ accounts for the fact that when new cards enter the system the values of $u$ get shifted to the left to make room for the new maximum possible value of tally marks. Since $x = k/n$ then if $n\to n+1$ larger value of $x$ are naturally shifted more than smaller values.   

If $v(x,\tau)$ denotes the continuous analogue of $P$, then the right hand side of the difference equation $–( P_t(k)-P_t(k-1))$ naturally gets transformed into $v_x(x)$. Using  Eqn. \ref{recursiveProb} for $P(k+1)/P(k)$ leads to the conclusion that the right hand of the difference equation is: 
\begin{equation}\label{vPrime}
v_x(x,\tau) = v(x,\tau)\left[ \frac{u_x(x,\tau)}{u(x,\tau)}-\frac{1}{x-U(x,\tau) }\right]
\end{equation}

As shown in the appendix, the difference equation then leads naturally to the following PDE:
\begin{equation}\label{PDE}
u_\tau(x,\tau) - (\omega x)u_x(x,\tau) = v(x,\tau)\left[ \frac{u_x(x,\tau)}{u(x,\tau)}-\frac{1}{x-U(x,\tau) } \right] 
\end{equation}
where $U(x,\tau)= \int_0^x u(x,\tau)dx$

Thus we have a continuous model of the card tally mark distribution.  While this is a rather intimidating system of PDEs we can straightforwardly examine the only steady state solution of this system.  

\subsection{Examining the Steady State Solution}

Since the familiarity curve appears to be a stable limiting solution, a natural object to investigate is the steady state solution of eq. \ref{PDE}.  Setting the $\tau$ derivative of $u$ equal to zero implies that:
\begin{equation} \label{steady}
 (\omega x)u_x(x,\tau)= v_x(x,\tau).
 \end{equation}

This remains a daunting expression, but it reveals the steady state behavior of $v$, the probability distribution for drawing a card with $x$ tally marks.  Indeed, integrating by parts and using the fact that $v(0)=v(1)=\omega$ we get that:
$$ v(x)= \omega( x u(x) - U(x) + 1 ).$$
Combining this with eq. \ref{vPrime} gives $$v_x = \omega( x u(x) - U(x) + 1 ) \left[ \frac{u_x(x,\tau)}{u(x,\tau)}-\frac{1}{x-U(x,\tau) }\right] $$

Thus eq. \ref{steady} can be written entirely in terms of $u$ and $U$:
$$x u_x(x,\tau) = ( x u(x) - U(x) + 1)\left[ \frac{u_x(x,\tau)}{u(x,\tau)}-\frac{1}{x-U(x) }\right] $$
Using $u_x = U_{xx}$ gives:
\begin{equation}
U_{xx} = \frac{(xU_x-U+1)}{1-U}\frac{U_x}{x-U}. \label{contTime}
\end{equation}
Since $U(0)=0$ and $U(1) = 1$ this equation is singular at both endpoints in the domain. However, it can be numerically solved in the interior of the domain via a shooting method.  

\begin{figure*}
\includegraphics[width=.8 \linewidth]{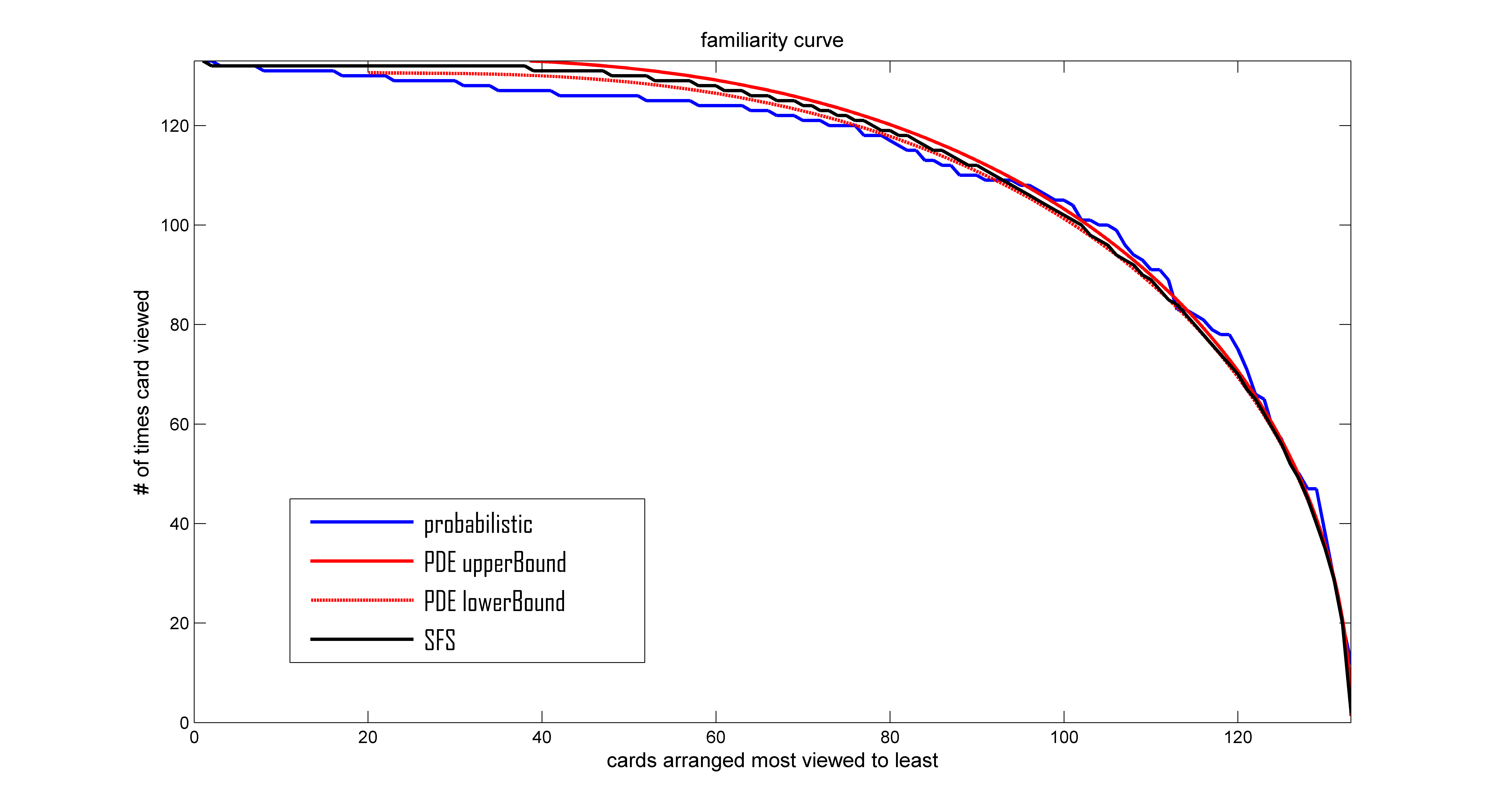} 
\caption{Numerically solving the steady state solution of the PDE model yields theoretical lower and upper bounds on the deterministic slow flashcard system.}\label{numInt}
\end{figure*}

Notice that as in Fig. \ref{numInt} the bounds predicted by Eqn. \ref{contTime} precisely describe the longterm numerical behavior of the SFS.  Interestingly, these bounds better predict the longterm behaviour of the SFS than the finite sized numerical simulations of the probabilistic flashcard system from which the PDE was derived.  Moverover, as shown in the next section, this PDE approach can be extended to similar flashcard systems, and explains the universality of familiarity curve.

\subsection{Faster Reinsertions Schemes}

\begin{figure*}
\includegraphics[width=1\linewidth]{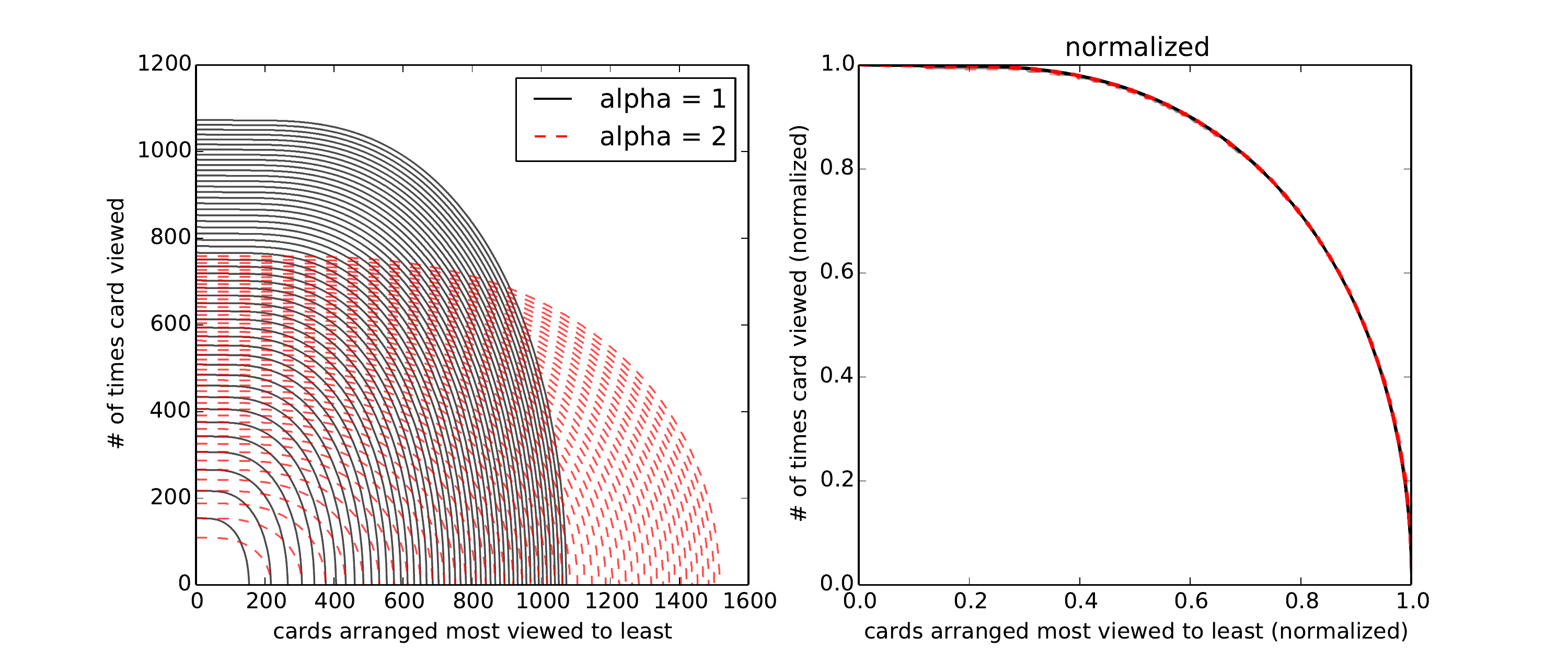} 
\caption{While faster reinsertion schemes produce different familiarity curves, they share the same normalized shapes. 
} \label{DifferentAlpha}
\end{figure*}

The slow flashcard system is the slowest of all reinsertion schemes where cards are inserted at strictly increasing depths.  A natural extension would consider inserting cards at position $r(k) = \alpha k +\beta$. In this setting let $\hat{k}(t)$ be the maximum number of tallymarks in the deck at some time $t$.  First, we will note that $\beta$ is unimportant.

For example, if $\beta = 2$, $r(k) = k+2$, then the card in position $\hat{k}(t)+2$ must have $\hat{k}(t)$ tallymarks on it. 
Exactly as before each time a card of maximal tallymarks is drawn, the number of active cards with $0$ tallymarks increases exactly by one.  Moreover, since each card with $k$ tallymarks now has $k+2$ positions it can be in, the probability that I draw a card with $k$ tallymarks, conditioned on not drawing a card with less than $k$ tallymarks is $\frac{f(k)}{k+2-g(k)}$.  While for a finite size system this expression is clearly different than Eqn. \ref{recursiveProbOne}, in the limit (as seen in the appendix) this effect disappears, as it does for all $\beta$. 

The more interesting case is when $\alpha >1$.  For example, consider $r(k) = \alpha k +\beta$. In this case, each time $\hat{k}$ increases, an additional $\alpha$ cards with $0$ tallymarks are inserted into the deck.  Similarly, since a card with $k$ tallymarks can be in $\alpha k +\beta$ different positions, the probability of drawing a card with $k$ tallymarks conditioned on not drawing a card with fewer tallymarks is $\frac{f(k)}{\alpha k +\beta-g(k)}$.   

However, despite what seems to be an extremely large difference in the rate that new cards are explored, and the speed at which a deck is explored, the scaled familiarity curve is the same as the slowflash card system Fig. \ref{DifferentAlpha}.  One way to understand the similarity of these two systems is to again consider the limit of the probabilistic version, as in the Appendix.  In particular, the continuous probabilistic version is independent of both $\alpha$ and $\beta$.  Interestingly, this suggests that the invariance of the familiarity curve is not a feature specific to the slow flashcard, but a fundamental feature of a family of systems, seemingly all at the cusp of the critical slowness in SFS. 

\section{Conclusion}

In this paper we have demonstrated that the combinatorial complexity of the SFS can be overcome through smoothing the complex determinism to a simpler probabilistic system.  Further, the probabilistic SFS and the associated PDE allow for the examination of related SFS, demonstrating that all linear insertion schemes achieve the same asymptotic familiarity curve, the universality of the familiar curve, and the self-critical behavior of linear reinsertion.  The universality of the familiarity curve suggests that there may be other fundamental archetypes of different learning schemes with large qualitative differences between them.  While this field remains young, determining the appropriate universality class for different settings may one day yield applicable and robust design criterion either for educational software or more generally.

\section{Acknowledgments}
This work was helped immensely through conversations with Tim Novikoff and Steven Strogatz.

\bibliography{Flashcards}

\section{Appendix}
Suppose cards are reinserted into position $r(k) = \alpha k+ \beta$.  In taking the continuum limit, consider the following change of variables:

\vspace{.1in}
\begin{tabular}{r l }

$x=$&$\mathrm{lim}_{t\to\infty} \frac{\alpha k}{n_t}$ \\ 
$\tau=$&$\mathrm{lim}_{t\to\infty} \int_0^t \frac{\alpha}{n_t} dt$ \\ 
\end{tabular} 
\vspace{.1in}

Where $n_t$ is the number of cards in the active deck at time $t$ and $x\in (0,\alpha]$.  Notice that since cards at reinserted at depth $r(k) = \alpha k+ \beta$, if the maximum tallymark in the deck is $\hat{k_t}$, then $n_t =  \alpha \hat{k}_t + \beta$.  Now we introduce the continuum limit of $f(k)$, $g(k)$ and $p(k)$.

\vspace{.1in}
\begin{tabular}{r l }
$u(x,\tau)=$&$ \mathrm{lim}_{k/n\to x} \alpha f_t(k)$ \\ 
$U(x,\tau)=$&$ \int_0^x u(\eta) d\eta=\mathrm{lim}_{k/n\to x} \frac{1}{\alpha n}  \sum_0^k \alpha f(j) =\mathrm{lim}_{k/n\to x} \frac{g(k)}{n}$ \\ 
$v(x,\tau)=$&$\mathrm{lim}_{n,k\to\infty} \alpha P(k)$.\\ 
\end{tabular} 
\vspace{.1in}

Correspondingly: 
\begin{eqnarray*}
u_x(x,\tau) &=& \mathrm{lim}_{h\to0} \frac{u(x+h)-u(x)}{h} \\
&=& \mathrm{lim}_{k/n\to x} \frac{ \alpha f(k+1)-\alpha f(k)}{\frac{\alpha}{n}} \\
&=& \mathrm{lim}_{k/n\to x} \frac{ f(k+1)- f(k)}{\frac{1}{n}}
\end{eqnarray*}

\begin{eqnarray*}
v_x(x,\tau) &=& \mathrm{lim}_{h\to0} \frac{v(x+h)-v(x)}{h} \\
&=& \mathrm{lim}_{k/n\to x} \frac{ P(k+1)-P(k)}{\frac{1}{n}} 
\end{eqnarray*}

\begin{widetext}
Now consider the limiting version of Eqn \ref{DiscreteEqn}, 

\begin{eqnarray*}
0 &=& \mathrm{lim}_{k/n\to x}  n_t  [-f_{t+1}(k)+
 f_t(k)+P(k-1)-P(k)] \\
& = & \mathrm{lim}_{k/n\to x} \frac{n_t}{\alpha} [-u(\frac{k_{t+1}}{n_{t+1}},\tau+\frac{\alpha}{n_t})+ u(\frac{k_{t+1}}{n_{t+1}},\tau) -u(\frac{k_{t+1}}{n_{t+1}},\tau)+
 u(\frac{k_t}{n_t},\tau)+  \alpha(P(k-1)-P(k))]
\end{eqnarray*}

Where the term $u(\frac{k_{t+1}}{n_{t+1}},\tau)$ was added and subtracted to complete what will be the derivative terms, $u_x$ and $u_\tau$.  

We now analyze the limits of each of these terms separately.  

Notice that $$\mathrm{lim}_{k/n\to x} \frac{-u(\frac{k_{t+1}}{n_{t+1}},\tau+\frac{\alpha}{n_t})+ u(\frac{k_{t+1}}{n_{t+1}},\tau)}{\frac{\alpha}{n_t}}=-u_\tau(x,\tau)$$

\begin{eqnarray*}
 \mathrm{lim}_{k/n\to x} \frac{ -u(\frac{k_{t+1}}{n_{t+1}},\tau)+
 u(\frac{k_t}{n_t},\tau)  }{\frac{\alpha}{n_t}}
&=&
u_x(x,\tau) \mathrm{lim}_{k/n\to x} \frac{\frac{k}{n_{t}}-\frac{k}{n_{t+1}} }{\frac{\alpha}{n_t}}\\
&=& u_x(x,\tau) \mathrm{lim}_{k/n\to x} \frac{\frac{k}{n_{t}}-\frac{k}{n_{t}+\alpha \omega(t)} }{\frac{\alpha}{n_t}}\\
&=& (x\omega) u_x(x,\tau) 
\end{eqnarray*}

Thus:
$$ u_\tau(x,\tau) - (\omega x)u_x(x,\tau)= - v_x(x,\tau)$$

However, we can also understand $v_x$ as:
\begin{eqnarray*}
& & v_x(x,\tau)=\mathrm{lim}_{k/n\to x} \frac{\alpha(P(k+1)-P(k))}{\frac{\alpha}{n}} \\
&=&v(x,\tau) \mathrm{lim}_{k/n\to x} \frac{1}{\frac{\alpha}{n}} \left[ \frac{\alpha k + \beta-g(k+1)}{\alpha k+\alpha+\beta-g(k+1)}\frac{f(k+1)}{f(k)}-1 \right] \\
&=&v(x,\tau)\mathrm{lim}_{k/n\to x} 
\frac{\alpha k + \beta-g(k+1)}{\alpha k+ \alpha +\beta -g(k+1)}\frac{f(k+1)-f(k)}{ \frac{1}{n}}\frac{1}{\alpha f(k)} 
 -\frac{\alpha}{ \frac{\alpha}{n}(\alpha k +\alpha +\beta -g(k)) } \\
&=&v(x,\tau)\left[ \frac{u_x(x,\tau) }{u(x,\tau)}-\frac{1}{x-U(x,\tau) }\right]
\end{eqnarray*}
\end{widetext}

We now take this PDE and use it to find a stationary solution. Setting the $\tau$ derivative equal to zero implies that:
$$ (\omega x)u_x(x,\tau)= v_x(x,\tau).$$
In the new scenario, it's important to consider briefly the boundary conditions.  Notice that each time a card of maximum tallymarks is drawn, the size of the active deck increases by $\alpha$.  Similarly, each time a card of $\hat{k}-1$ is drawn the number of active cards available for advancement by cards with $\hat{k}-1$ tallymarks increasings by exactly one.  Since $\mathrm{lim}_{k/n\to x} P(\hat{k}-1)=v(1)=\mathrm{lim}_{k/n\to x} P(\hat{k})$ in the limit then $v(0) = \omega = v(1) $.

We now take $v_x = \omega x u_x$ and integrate by parts:
\begin{eqnarray*}
\int_0^x v_x(\eta) d\eta &=& \int_0^x  \omega \eta u_x(\eta) d\eta \\
v(x)-\omega &=& \omega(ux-U(x)) \\
v(x) &=& \omega(xu(x)-U(x)+1)
\end{eqnarray*}

We now use equate our two expressions for $v_x(x)$ and substitute in our the above expression for $v(x)$ giving:

\begin{eqnarray*}
(\omega x) u_x(x)&=& \omega(xu(x)-U(x)+1)\left[ \frac{u_x(x,\tau) }{u(x,\tau)}-\frac{1}{x-U(x,\tau) }\right]
\end{eqnarray*}

Since $u_x = U_{xx}$ and $u=U_x$ this gives:
\begin{equation}
U_{xx} = \frac{(xU_x-U+1)}{U-1}\frac{U_x}{x-U} \label{Uxx}
\end{equation}
Notice that neither $\alpha$ nor $\beta$ are present in Eqn. \ref{Uxx} suggesting that there is the same universal familiarity curve for all such variants of the SFS.

\end{document}